\documentclass[reprint,superscriptaddress, floatfix,aps,prd]{revtex4-2}
\usepackage[utf8]{inputenc}
\usepackage{amsfonts}
\usepackage{amssymb}
\usepackage{amsmath}
\usepackage{csquotes}
\usepackage{bbm}
\usepackage{verbatim}
\usepackage{graphicx}
\usepackage{xcolor}
\usepackage{hyperref}
\usepackage{listings}
\usepackage{enumerate}
\hypersetup{
    colorlinks = true,
    linkcolor = red,
    anchorcolor = blue,
    citecolor = blue,
    filecolor = blue,
    urlcolor = blue
    }

\usepackage{setspace}

\begin{document}

\preprint{APS/123-QED}

\title{Exploring Lee-Yang and Fisher Zeros in the 2D Ising model through multipoint Padé approximants}

\author{Simran Singh}
 \email{ssingh@physik.uni-bielefeld.de}
\affiliation{Universität Bielefeld, Fakultät für Physik, D-33615 Bielefeld, Germany}
\author{Massimo Cipressi}
\email{massimo.cipressi@studenti.unipr.it}
\affiliation{Dipartimento di Scienze Matematiche, Fisiche e Informatiche, Università di Parma, I-43100 Parma, Italy}
\author{Francesco Di Renzo}
\email{francesco.direnzo@unipr.it}
\affiliation{Dipartimento di Scienze Matematiche, Fisiche e Informatiche, Università di Parma and INFN, Gruppo Collegato di Parma
I-43100 Parma, Italy}

\graphicspath{{images/}}

\date{\today}

\begin{abstract}
We present a numerical calculation of the Lee-Yang and Fisher zeros of the 2D Ising model using multi-point Pad\'{e} approximants. We perform simulations for the 2D Ising model with ferromagnetic couplings both in the absence and in the presence of a magnetic field using a cluster spin-flip algorithm. We show that it is possible to extract genuine signature of Lee Yang and Fisher zeros of the theory through the poles of magnetization and specific heat, using multi-point Pad\'{e} method.  We extract the poles of magnetization using Pad\'{e} approximants and compare their scaling with known results. We verify the circle theorem associated to the well known behaviour of Lee Yang zeros. We present our finite volume scaling analysis of the zeros done at $T=T_c$ for a few lattice sizes, extracting to a good precision the (combination of) critical exponents $\beta \delta$. The computation at the critical temperature is performed after the latter has been determined via the study of Fisher zeros, thus extracting both $\beta_c$ and the critical exponent $\nu$. Results already exist for extracting the critical exponents for the Ising model in 2 and 3 dimensions making use of Fisher and Lee Yang zeros. In this work, multi-point Pad\'{e} is shown to be competitive with this respect and thus a powerful tool to study phase transitions. 
\end{abstract}

\maketitle

\section{\label{sec:Intro} Introduction}
\noindent The knowledge of phase transitions is essential in determining the equation of state of a physical system. Phase transitions are characterised by non-analyticities that develop in thermodynamic functions in the limit of infinite degrees of freedom in a system. In finite volume systems, thermodynamic functions like the free energy are analytic functions of system parameters and the study of the onset of divergences becomes challenging. However, there are remnants of these divergences (if existing) even in finite volumes and occur as peaks in the susceptibility of the order parameter, whose height and width scale with volume. The conventional way to look for phase transitions in finite systems is to study the finite size scaling of these susceptibilities \cite{Landau:1976zz, ADBruce_1981, Binder:1981zz, Binder:1981sa, Binder:1984llk,Engels:2014bra}. Although these techniques are still used extensively in the study of phase transitions in the fields of condensed matter and lattice gauge theories, one of our goals in this paper is to show the power of an alternative scheme in extracting critical exponents for certain theories from their numerical simulations at finite volumes, namely, the Lee-Yang (LY) zero analysis \cite{LeeYang1952I,LeeYang1952II,Kortman:1971zz,Fisher:1978pf,Itzykson:1983gb}. While results on the numerical extraction of LY zeros from some lattice models already exist \cite{Marchesini:1988hv,Matveev:1994wg,Matveev:1994ha,Matveev:1995fa,Matveev:1995gh,PhysRevLett.81.5644,PhysRevLett.118.180601,PhysRevE.96.032116,IoanaBenaLY17,PhysRevE.97.012115,Deger:2019mgo,MAJUMDAR2020124263,PhysRevA.100.022125}, the novelty of our work is to extract them using only the leading order cumulants of the partition function, evaluated at multiple points in parameter space,  using a rational function re-summation of the cumulants called the \enquote{multi-point Pad\'{e}} method.

\noindent Another motivation for this work comes from our recent studies of the QCD phase diagram in the complex chemical potential plane. We have shown \cite{Dimopoulos:2021vrk} that it is possible to gain more information from the generated Taylor coefficients, if we re-sum them into a rational function. In the present paper, we aim at providing confidence in the claim that the multi-point Pad\'e method can extract genuine LY zeros, illustrating it in the 2D Ising model.
The Ising model is a simple physical system displaying a phase transition in dimension $D\geq 2$ . Although an exact solution of the 2D Ising model exists due to Onsager in the thermodynamic limit, numerical simulations usually have to be performed for any finite size results. These have to be {\em a fortiori} accurate if we are interested in a numerical study of divergences arising in the thermodynamic limit. These simulations are based on Monte Carlo methods, and hence are plagued by statistical errors. The presence of such errors is one of the reasons we want to apply the multi-point Pad\'{e} analysis to study the LY zeros; we want to check whether the tool is robust enough to live with finite accuracy of input data. A preliminary analysis leading up to this work can be found in \cite{DiRenzo:2023xeg,SS:2022thesis}.

\noindent We will begin our analysis by giving an overview of the complex zeros of the partition function as applied to the Ising model, putting emphasis on the well known properties of the LY zeros in Section~\ref{sec:LYZ}. In Section~\ref{sec:Ising} we will briefly describe the simulation procedure used for the 2D Ising model and outline the phase transitions we are looking for.
In Section~\ref{sec:Pade}, we will give a brief overview of the multi-point Pad\'e method and the error analysis procedure used. In Section~\ref{sec:ScalingLY} we will describe results for the scaling analyses used. In Section~\ref{sec:Conclusions} we will conclude with some results and outlook.

\section{\label{sec:LYZ}Complex zeroes of the partition function}
\noindent We begin by explaining the origin and nature of complex zeros of the partition function, in the context of the Ising model in the presence of an external magnetic field $H$.
The zeros of the partition function in the complex $H$ plane are known as the Lee-Yang (or Yang-Lee) zeros \cite{LeeYang1952I,LeeYang1952II}. We can also discuss the zeros of the partition function in the absence of an external magnetic field, called the Fisher zeros \cite{Fisher:1978pf}, which appear in the complex inverse temperature $\beta$ plane. However, some of the properties shown by these zeros, like the \emph{Circle theorem} \cite{LeeYang1952I,LeeYang1952II} only hold for zeros in the complex $H$ plane, while other properties are common to both kinds of zeros. We will focus our attention on LY zeros in the following. We further restrict our study to nearest neighbour ferromagnetic interaction. The Hamiltonian describing this theory is given by (with $J > 0$)
\begin{equation}\label{eqn:hamil}
    \mathcal{H} = - J \sum_{\langle i j \rangle} \sigma_i \sigma_j - H \sum_i \sigma_i \,\,\, ,
\end{equation}
where $J$ is the coupling and $\sigma_i \in \{\pm1\}$ are the spins at the site $i$. The canonical partition function is given by:
\begin{equation}\label{eqn:partn}
    Z_N = \sum_{\sigma_i \in \{-1,1\}} e^{-\beta \mathcal{H}}
\end{equation}
\noindent From the form of the Hamiltonian, it can be seen that up to an overall functional dependence on $z = e^{\beta H}$, the partition function is a polynomial in $e^{\beta H}$ of order $2 N$, with $N$ being the number of sites on the lattice \footnote{This is equivalent to the statement that the partition function is a polynomial in $e^{2 \beta H}$ of order $N$.}. Therefore, the order of the polynomial grows linearly with the number of sites. This remark will become important in the discussion to follow on the relation of LY and Fisher zeros to phase transitions. 

\noindent In order to see why the partition function is a polynomial, consider the term $e^{\beta H \sum_i \sigma_i}$. For $N$ lattice sites, the partition function will be a sum over $2^N$ terms weighted by the factor $e^{\beta J \sum_{\langle i j \rangle} \sigma_i \sigma_j}$ that depends on the spin configuration. Note that this factor is invariant under the transformation of flipping all the spins on the lattice. Additionally, there will be a permutation factor associated with the number of spins pointing up or down which will be the same for configurations related by spin flips \footnote{This factor takes into account the number of ways $m$ up spins and $N-m$ down spins can be put on a lattice, given by $\frac{N !}{m! (N-m)!}$}. Taking these factors into account, the partition function can be written as:

\begin{align}
 Z_N(z) &= a*z^N + b*z^{N-2} + c*z^{N-4} + \hdots \nonumber \\
 &+c*z^{-N+4} + b*z^{-N+2} + a*z^{-N} \nonumber \\     
 &= z^{-N} * \left(a + b*z^{2} + c*z^{4} + \hdots \right. \nonumber \\
 &\left. + c*z^{2 N-4} + b*z^{2 N-2} + a*z^{2 N} \right)
 \label{eq:partNF}
\end{align}

The symmetry of the partition function under $z \to z^{-1}$ is the statement that the Hamiltonian in Eq.~(\ref{eqn:hamil}) is invariant under the combined action of reversing the external magnetic field and flipping all the spins.

\noindent A few comments can be made based on Eq.~(\ref{eq:partNF}) above :
\begin{enumerate}[I]
    \item The partition function has the functional form of an even polynomial multiplied by a factor $z^{-N}$. We could factor the partition function this way because $z$ can never be zero. 
    \item For real ($\beta,H$), the coefficients of the polynomial are strictly positive. This implies that complex roots always occur in complex conjugate pairs, in the complex $H$ plane, for fixed, real $\beta$ and in the complex $\beta$ plane for fixed, real $H$.
    \item As the lattice volume $N$ increases, the order of the polynomial increases, which leads to an increase in the number of complex zeros of $Z_N$. In the limit of $N \to \infty$, these zeros accumulate and coalesce into cuts. A phase transition does not occur when the zeros do not approach the real axis when increasing $N$.
    \item A phase transition is said to occur at some critical temperature $T_{crit}$ when these zeros approach the real axis of the external field parameter, in the thermodynamic limit. The behaviour of the density of (Lee-Yang) zeros at the real axis is used to distinguish between a first and a second order phase transition. At a second order transition the complex conjugate pair gets infinitesimally close to the real axis but the density of zeros is zero on the real axis. On the other hand, a first order transition sees a non-zero density of zeros on the real axis. 
    \item Furthermore, because of the even nature of the polynomial, if $z$ is a root, then so is $-z$.
\end{enumerate}

\noindent Therefore, for a finite $N$, the partition function is strictly positive when $\beta$ and $H$ are real. Hence, the only zeros of $Z_N$ occur in the complex plane of $z$ or on the $z<0$ axis. In the complex $H$ plane this translates to having \emph{only complex} zeros of $Z_N(z(H))$. However, in the infinite volume limit, $Z_N$ is an infinite series which can now have non-trivial zeros on the $z>0$ axis, or real zeros in $H$. Since real zeros of the partition function mark the onset of phase transitions, we have recovered the well-known result that phase transitions cannot occur in a finite volume.  The natural question to ask now will be on how to access these complex zeros of the partition function. This is fortunately not hard to answer because thermal cumulants, like the average magnetization in the case of the Ising model, are related to the derivative of the logarithm of the partition function (see Section~\ref{sec:ScalingLY}). This means that the zeros of $Z_N(H)$ will appear as \emph{poles} of the average magnetization as a function of $H$ and the specific heat capacity as a function of $\beta$. The aim of the next few sections is to study in some detail these poles in the complex plane of the external magnetic field and inverse temperature. Arguments similar to those presented in this section can also be found in more detail in the existing literature \cite{Itzykson:1983gb,Matveev:2008}, and we refer the reader to \cite{IoanaBenaLY17} for a detailed review on the subject.

\section{\label{sec:Ising}Simulating 2D Ising model}

\noindent The Ising model has been extensively studied in the literature and can also be found in many textbooks \cite{newmanb99,Thijssen2007,entropy} on statistical mechanics. The model has an exact solution in 1D due to Ising \cite{Ising:429052} and does not undergo any phase transition. However, its 2D version is one of the simplest systems to undergo a continuous (or $2^{nd}$ order) phase transition from a symmetry broken (ferromagnetic) phase to a symmetric (paramagnetic) phase at $T_c = \frac{2 J}{\ln{(1+\sqrt{2})}} \sim 2.269 \,J$. The model can also be seen to undergo a discontinuous (or $1^{st}$ order) phase transition when considering the average magnetization (Eq. \ref{eq:avMag}), as a function of $H$ at $T \leq T_c$, across $H=0$. An exact solution for this model in the thermodynamic limit exists due to Onsager \cite{Onsager1944}. Hence we know the transition temperature and the various critical exponents characterizing this transition analytically. Because of this, the 2D Ising model serves as an ideal candidate for testing new numerical methods. In general, extracting critical exponents from numerical data is a non-trivial task and requires formidable amounts of statistics. Thankfully, the 2D Ising model is easy to simulate and not expensive where getting large statistics is concerned. Hence, we choose to test our multi-point Pad\'{e} method on this model. \\
\noindent The 2D Ising model can be simulated using both single spin flip \footnote{for a detailed analysis of extracting the critical exponents from single spin flip algorithms see \cite{Ibarra-Garcia-Padilla:2016cox}} and cluster spin flip algorithms\cite{Swendsen:1987ce,Wolff:1988uh}. It is well known that single spin flip algorithms suffer from critical slowing down \cite{Wolff:1989wq}, and since our goal is to extract LY zeros close to and at the critical temperature $T_c$, we will use a cluster spin flip algorithm based on \cite{DESTRI1992311}, where a modification to the original Swendsen-Wang algorithm \cite{Swendsen:1987ce} was described, to add an external magnetic field parameter in the code. We will discuss below two types of simulations performed. Note that the ferromagnetic coupling constant $J$ has been set to $1$ for the simulations that follow. 

\subsection{Simulation : To extract Fisher zeros}

\noindent For the first part of our analysis, we want to study the \emph{Fisher zeros} - zeros of the partition function in the complex $\beta$ plane. For this we compute the energy density $\langle E \rangle$ and the specific heat capacity $C_H$ of the 2D Ising model, with $H$ set to $0$. These quantities are given by
\begin{align}
    \langle E \rangle &= \frac{\partial \ln Z_N}{\partial (-\beta)} \nonumber \\
    C_H &= \left( \frac{\partial \langle E \rangle }{\partial T} \right)_H = \beta^2 \left(\langle E^2 \rangle - \langle E \rangle^2 \right)
\end{align}
with $E = - J \sum_{\langle i,j \rangle} \sigma_i \sigma_j$. The model is simulated at temperature values given by $T \in \{1.76, \hdots 2.15\}$ in steps of $0.03$, $\{2.17, \hdots 2.40 \}$ in steps of $0.01$ and $\{2.43, \hdots 3.00\}$ in steps of $0.03$ with $H=0$. These are then repeated for different lattice volumes at $L \in \{10,20,40,64,80 \}$, to perform the finite volume analysis of the Fisher zeros obtained. 

\begin{figure}
    \centering
    \includegraphics[width=.485\textwidth]{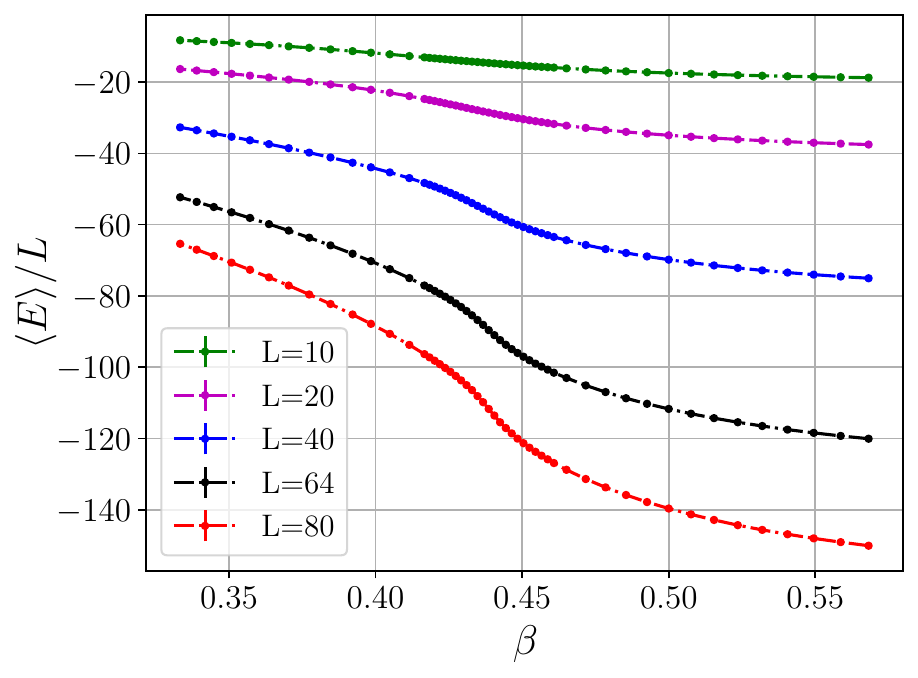}
    \includegraphics[width=.475\textwidth]{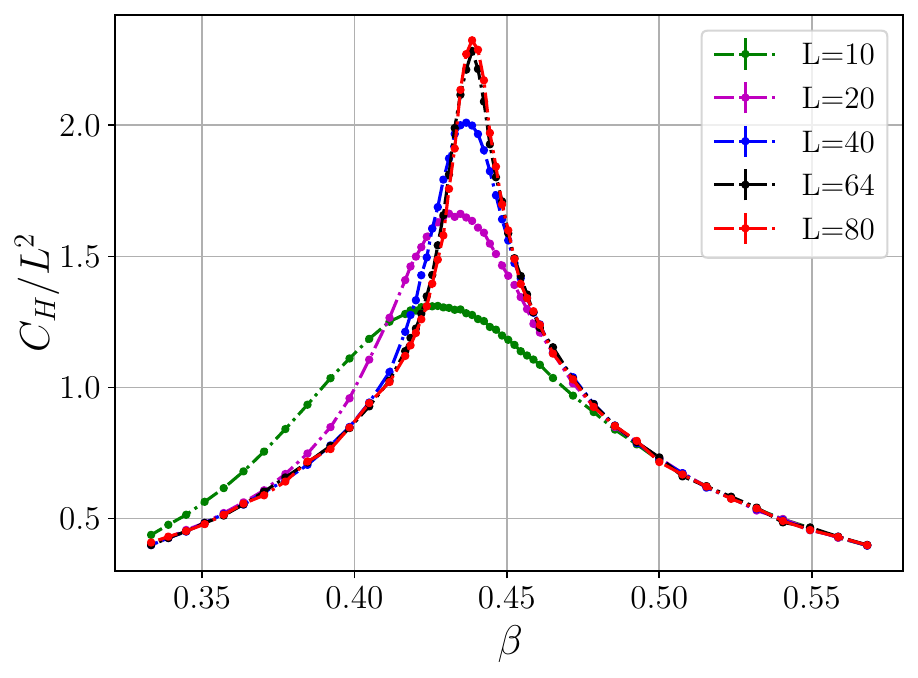}
    \caption{Top : Average energy (scaled by $1/L$ instead of $1/L^2$ for better visualization) as a function of $\beta$ from simulations. Bottom : Specific heat capacity per lattice site calculated from the configurations generated. }
    \label{fig:sim_details2}
\end{figure}

\noindent We refer the reader to Fig.~\ref{fig:sim_details2} for the results of the simulation for $\langle E \rangle$ and $C_H$. As an indication for the kind of statistics used, we list here the number of configurations used per lattice size. For $L \in \{10,20,40,64,80\}$, the following number of configurations were used respectively: $\{300K,125K,125K,40K,25K \}$. Before analysing this data, we detail the simulations required for the LY zero analysis. 


\subsection{Simulation : To extract Lee-Yang zeros} 

\begin{figure}
    \centering
    \includegraphics[width=.485\textwidth]{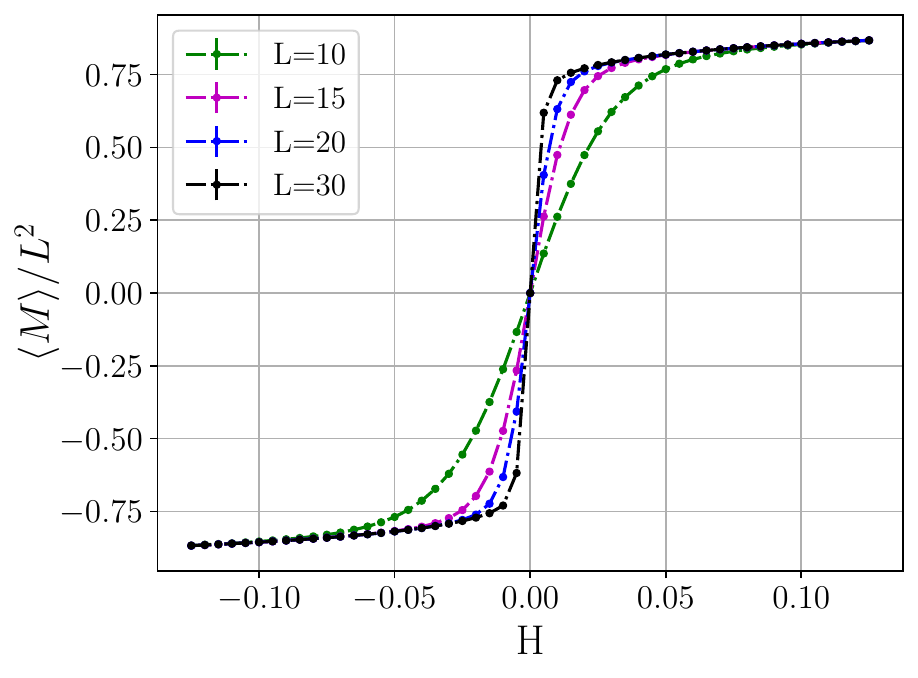}
    \includegraphics[width=.475\textwidth]{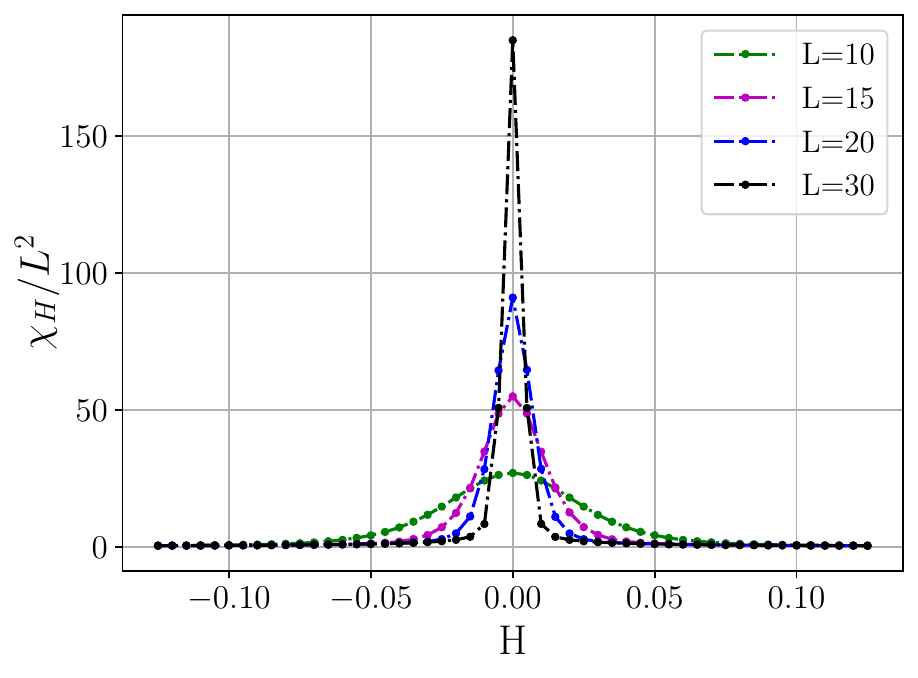}
    \caption{Top : Average magnetization per site as a function of $H$ from simulations. Bottom : Susceptibility per site calculated from the configurations generated. }
    \label{fig:sim_details1}
\end{figure}

\noindent For the second part of the analysis, we want to study the zeros of the partition function in the complex $H$ plane.
For this we need to compute the average magnetization $\langle M \rangle$ and susceptibility $\chi_H$ of the model, given by
\begin{align}
\label{eq:avMag}
    \langle M \rangle &= \frac{1}{\beta} \frac{\partial \ln Z_N}{\partial H} \nonumber \\ 
    \chi_H &= \left( \frac{\partial  \langle M \rangle}{\partial H} \right)_T = \beta \left(\langle M^2 \rangle - \langle M \rangle^2 \right) 
\end{align}
with $M = \sum_i \sigma_i$. In order to verify the volume scaling of the LY zeros, we simulate lattice volumes of sizes $L \in \{10,15,20,30\}$. Each lattice volume was simulated for $H \in \{-0.125, \hdots 0.125 \}$, in steps of $0.005$ at $T=T_c \sim 2.269 \, J$. \\ \noindent The resulting $\langle M \rangle$ and $\chi_H$, per lattice site, are shown in Fig.~\ref{fig:sim_details1} and the details of the number of configurations are as follows: For each lattice size, $625K$ configurations were used to estimate the average magnetization and the resulting susceptibility. We will now proceed to describe the multi-point Pad\'e method for extracting the zeros of the partition function from the poles of the $C_H$ and $\langle M \rangle$ data.

\section{\label{sec:Pade}Multi-point Pad\'{e} method}

\noindent Pad\'{e}-type rational approximations have recently (re-) emerged as a reliable tool to re-sum Taylor series coefficients in the studies of lattice QCD \cite{DiRenzo:2020cgp,Pasztor:2020dur,Basar:2021gyi,Costin:2021bay,Dimopoulos:2021vrk,Bollweg:2022rps}. Most of the literature that exists toward the existence, convergence and uniqueness of solutions for Pad\'{e} approximations is limited to only a restricted class of functions to be approximated \cite{baker1975,Fraser1962OnTC,MontessusDeBallore,NUTTALL1970147,POMMERENKE1973775,CUYT1996141,Nuttal2Rieman,Gruyter}. Instead, many of the interesting results on Pad\'{e} approximants are known only due to numerical experiments like those in \cite{yamada2014numerical,Masjuan:2010ck,Costin:2022hgc,Basar:2021gyi}. However, most of the literature referred to above is based on what is called the \enquote{single point Pad\'{e}} approximation. This involves using Taylor series coefficients of the unknown function about a single point and constructing a rational approximation using these coefficients. An immediate limitation of this method is the need to have relatively high number of Taylor coefficients to build rational functions of increasing order. Numerical simulations typically do not allow for the generation of such high order Taylor coefficients with reasonable accuracy. One can instead use lower order Taylor coefficients of the function evaluated at \emph{multiple} points, which forms the basis of our analyis. This trade-off allows us to construct high order rational approximants. In the following we will only focus on the construction of such approximants and briefly outline the sources of error. For a more detailed discussion on this method and results of numerical experiments, we refer the reader to \cite{Dimopoulos:2021vrk}.
\subsection{The multi-point Pad\'e method in a nutshell}
\noindent Let us consider the rational function $R^{m}_{n}(z)$
\begin{equation}
\label{eq:PadeRatFunct}
R^{m}_{n}(z) = \frac{P_m(z)}{\tilde{Q}_n(z)} = \frac{P_m(z)}{1+Q_n(z)} = \frac{\sum\limits_{i=0}^m \, a_i \, z^i}{1 + \sum\limits_{j=1}^n \, b_j \, z^j}\,
\end{equation}
with $m$ and $n$ being the degrees of the polynomials at numerator and denominator respectively.
Writing $\tilde{Q}_n(z) = 1+Q_n(z)$ ensures that the rational function depends essentially on $n+m+1$ parameters. 
Notice that one should naturally also demand that there is no point $z_0$ such that
$ P_m(z_0) = \tilde{Q}_n(z_0) = 0$, \emph{i.e.} we should in principle exclude any (common) zero of both
numerator and denominator. If this were not the case, we would have rather essentially defined the rational function
$R^{m'}_{n'}(z)$ with $n=n'+l$ and $m=m'+l$ for some integer $l>0$. Despite this, we will nevertheless not exclude the possibility of common zeros, for a reason that will be clear when we explain why we are interested in $R^{m}_{n}(z)$. \\
Let us consider now a function $f(z)$ and suppose we know a few of its Taylor expansion coefficients,
let's say at different points $\{z_k\,|\,k=1 \ldots N\}$. The number of coefficients we know can be different
at different points, but for the sake of simplicity we will assume that $f^{(s-1)}$ is the highest order derivative
which is known at each point, together with all derivatives of degree $0 \leq g < s-1$. Then the system of equations we have to solve becomes

\begin{equation}
\label{eq:LinearProblem3}
\begin{split}
P_m(z_k) - f(z_k)Q_n(z_k) &= f(z_k) \\ 
P_m'(z_k) - f'(z_k)Q_n(z_k) - f(z_k)Q_n'(z_k) &= f'(z_k) \\ 
 & \vdots \\
P_m^{(s-1)}(z_k) - f^{(s-1)}(z_k)Q_n(z_k) - \hdots &-f(z_k)Q_n^{(s-1)}(z_k)\\
&= f^{(s-1)}(z_k) \\
\end{split}
\end{equation}

\noindent In what follows, we will solve this system of linear equations to determine the coefficients of the polynomials $P_m$ and $Q_n$ while restricting our analysis to diagonal ($[q,q]$) and near diagonal ($[q,q+1]$) type Pad\'e approximants. Other techniques like a generalized $\chi^2$ method can also be used to estimate the coefficients of the rational functions by minimizing the distance between the measured Taylor coefficients and the required rational function, weighted by the estimated errors on the measured coefficients. This has been compared to the linear solver method in \cite{Dimopoulos:2021vrk}.

\subsection{The method at work for the 2D Ising model}

\begin{figure}
	\centering
	\includegraphics[width=.5\textwidth]{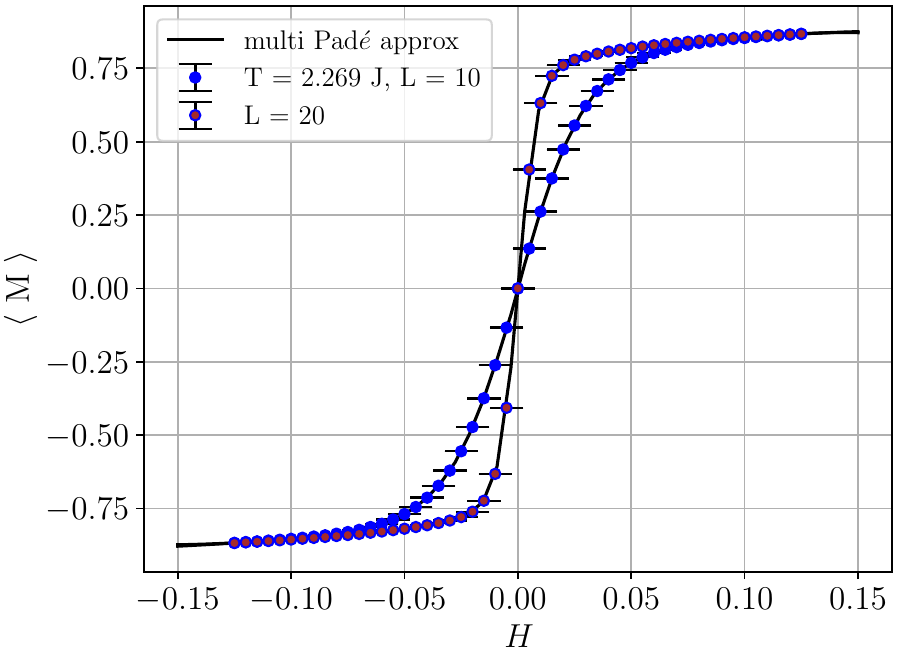}
	\hfill
	\includegraphics[width=.5\textwidth]{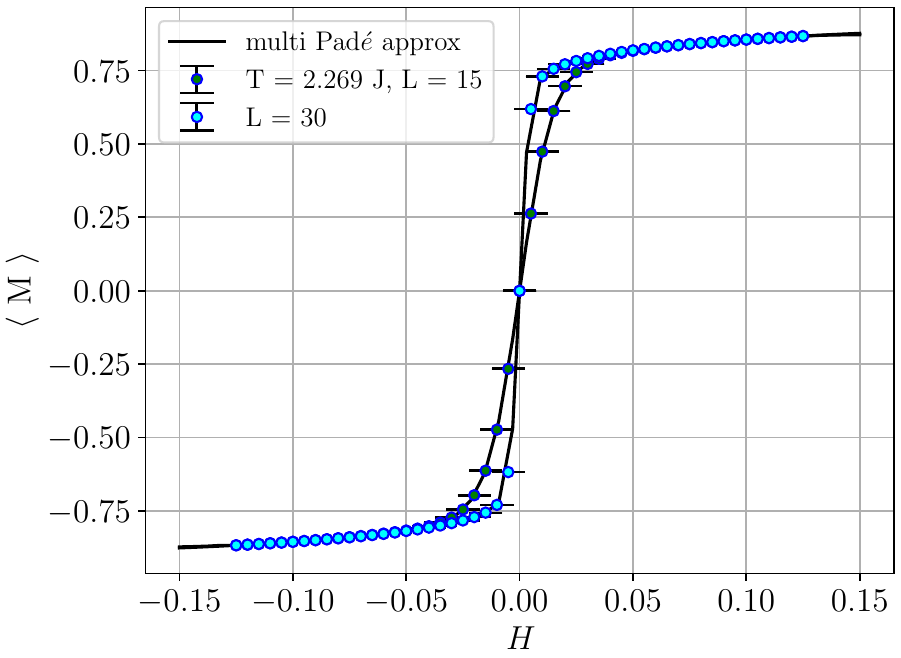}
	\caption{Average magnetization as a function of the external magnetic field, rational function approximation vs data. Top : $L = 15,20$ and Bottom : $L = 10,30$ (Plotted separately for sake of clarity). The rational approximation shown has the order $[m,n] = [25,25]$.}
        \label{fig:RatApprox}
\end{figure}

\begin{figure}
	\centering
	\includegraphics[width=.5\textwidth]{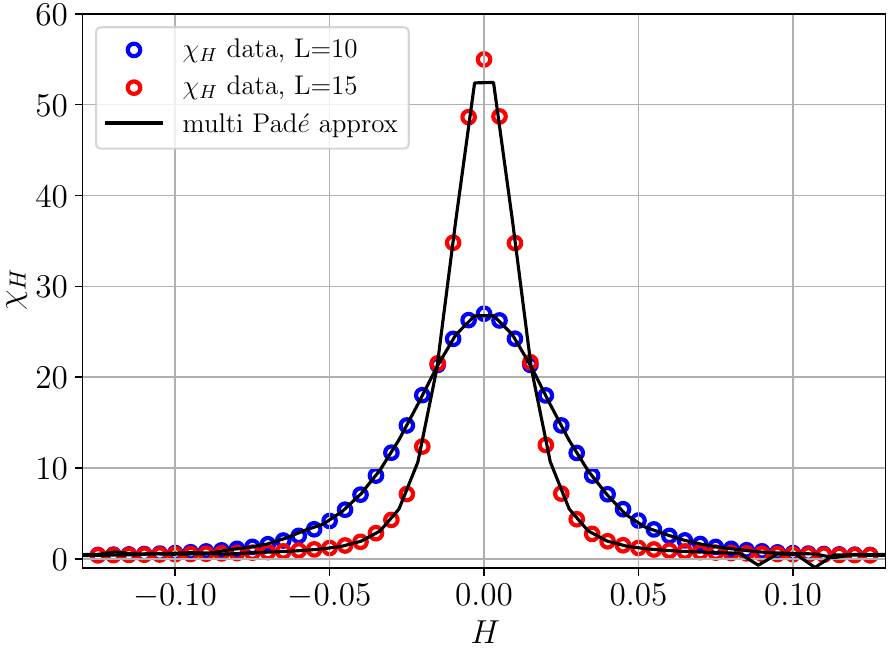}
\caption{Derivative of the rational function obtained in Fig.~\ref{fig:RatApprox} for $L=10 \,\,, 15$ plotted against the susceptibility data. Note that this is not an interpolation and no data on the susceptibility was used in the construction of the rational function. Since this is a derivative of an $[m,n]=[25,25]$ rational function, the order is $[m,n]=[24,25]$}
        \label{fig:RatApproxChi}
\end{figure}

\noindent We will now focus on the results of the approximation and the singularity structure obtained from the Pad\'{e} procedure outlined in the previous sub-section, for the average magnetization. We will first show the results of the approximation in Fig.~\ref{fig:RatApprox}. That we can see the rational functions approximate the data correctly is not surprising because we have essentially done a rational interpolation through the data, since our input only consisted of the average magnetization values and not the susceptibilities.
Therefore, the first real success of the rational approximation is to see how well its derivatives approximate the susceptibilities, as shown in Fig.~\ref{fig:RatApproxChi} for the $L=10 \,\,,15$ data. This is a nice result because the rational function was constructed assuming only the knowledge of the zeroth order Taylor coefficients in the expansion of the average magnetization and it faithfully returns the expected first order coefficients at all the input points \footnote{Notice that in Fig. 4, around $H\sim0.1$ we observe a small bump in the rational approximation of the susceptibility. This is due to a near cancellation of a zero-pole pair on the real $H$ axis (see Fig.~\ref{fig:LeeYangZerosL}, top). The reason these zero-pole pairs don't cancel exactly sometimes is because the data comes with noise. For a more detailed explanation see \cite{Dimopoulos:2021vrk}.}.
Having gained some confidence in our multi-point Pad\'{e} approximation, we can now proceed to look at the singularity structure, i.e. we will now study the zeros and poles of the rational function constructed in the complex $H$ plane \footnote{All results shown below have been obtained using the linear algebra routines of MATLAB \cite{MATLAB:2019} and all plots have been made using Jupyter notebooks \cite{soton403913}}.  

\noindent Selecting two lattice sizes at $T_c$, we refer the reader to Fig.~\ref{fig:LeeYangZerosL}.
In the figure, we depict the zeros of
  the numerator (black pentagons) and of the denominator (red crosses) of
  $R^{m}_{n}(H)$ at different values of the lattice size $L$, {\em
    i.e.} $L=15$ (left panel) and $L=30$ (right panel). The order the rational approximant constructed is $[m,n]=[25,25]$. The pale blue points are shown to indicate the interval of points where the value of average magnetization was used as an input to build the rational function. 
We can easily make a couple of key observations. 
\begin{itemize}
\item A few zeros of the denominator are canceled by corresponding zeros of
  the numerator. These are not genuine pieces of information: actually
  their location can vary when varying {\em e.g.} the order of the Pad\'{e}
  approximant $[m,n]$. On the other hand, genuine pieces of information ({\em i.e.}
  actual zeros and poles) stay stable to a very good
  precision. Notice that the alternating sequence of zeros and poles along the imaginary $H$ axis is how a Pad\'e approximant alludes to a branch cut eventually showing up in the thermodynamic limit \footnote{In principle we find other zeros and poles very far away from the region of interpolation, however these usually occur when we demand higher order of the Pad\'e approximant and are not stable feature of the approximated function. This kind of structure even occurs when we try to approximate analytic functions.}.
\item In order to analyse the genuine poles more carefully, we remove the \emph{spurious} poles from the singularity structure. We further remove all remaining zeros to focus the attention on poles and refer to Fig.~\ref{fig:cir_theorem}. Looking at the pole structure of the figures, we can make the following comments : {\em (1)} As the lattice size $L$ gets larger, the closest singularity $H_0$ gets closer to the real $H$ axis. We claim that we verify the circle theorem, which translates to the Lee-Yang zeros lying on the imaginary $H$ axis. However, looking at Fig.~\ref{fig:LeeYangZerosL} the reader may notice that some of the poles for the $L=30$ lattice are shifted away from imaginary $H$ axis, and hence \textit{seemingly} violating the $H \to -H$ symmetry. This is because we have used the central values of magnetization which were simulated at each quoted value of $H$ and hence do not respect the anti-symmetry exactly, but do so within errors. In support of our claim we provide the location of the genuine poles of magnetization at the central values along with error bars in Fig.~\ref{fig:cir_theorem}, obtained by performing a bootstrap procedure, i.e. for a given point we extract new data from a Gaussian distribution with mean given by central value and standard deviation given by the estimated error. Notice that only the closest pole tends to appear stable. We further observe that for the smaller $L=15$ lattice, the second pole has some uncertainty easy to inspect by eye. For the $L=30$, we observe a third pole which now appears with even larger uncertainty. In the following, only the closest pole is important for the discussions of scaling. {\em (2)} Referring the reader to Section \ref{sec:LYZ}, all poles occur with their complex conjugate pairs. {\em (3)} Although this observation is highly dependent on the statistics used, which for the purposes of this work is relatively high, we can see an increase in the number of zeros as the number of lattice sites increases. All of these points seem to indicate that we are observing genuine Lee-Yang zeros.
\end{itemize}

\subsection{On the stability of the closest poles to the real $H$ axis}

\begin{figure}
	\centering
	\includegraphics[width=.52\textwidth]{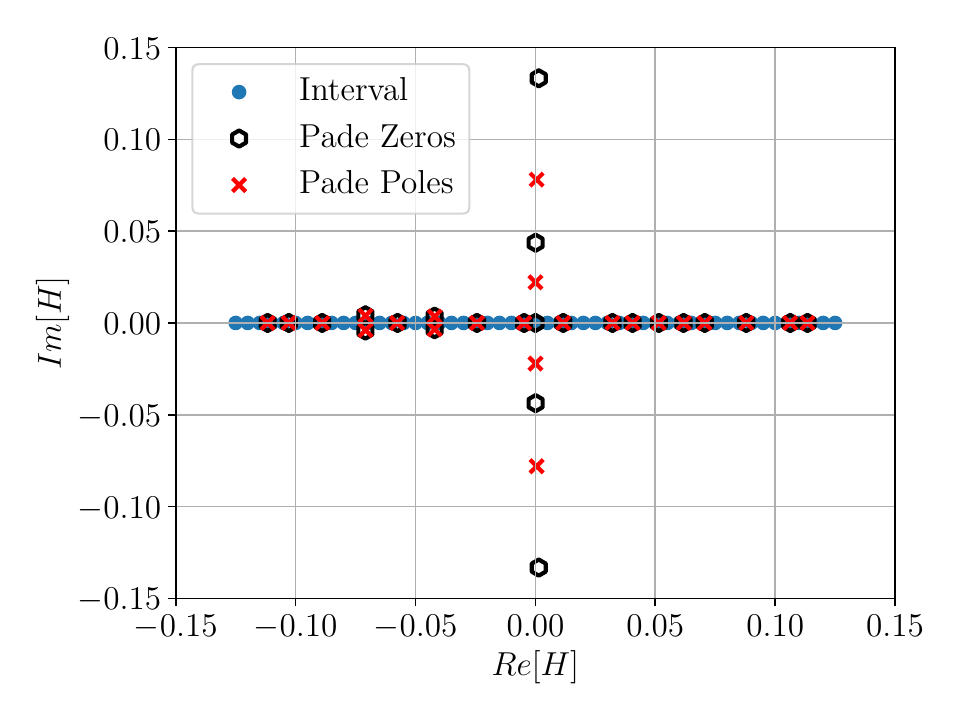}
	\hfill
	\includegraphics[width=.52\textwidth]{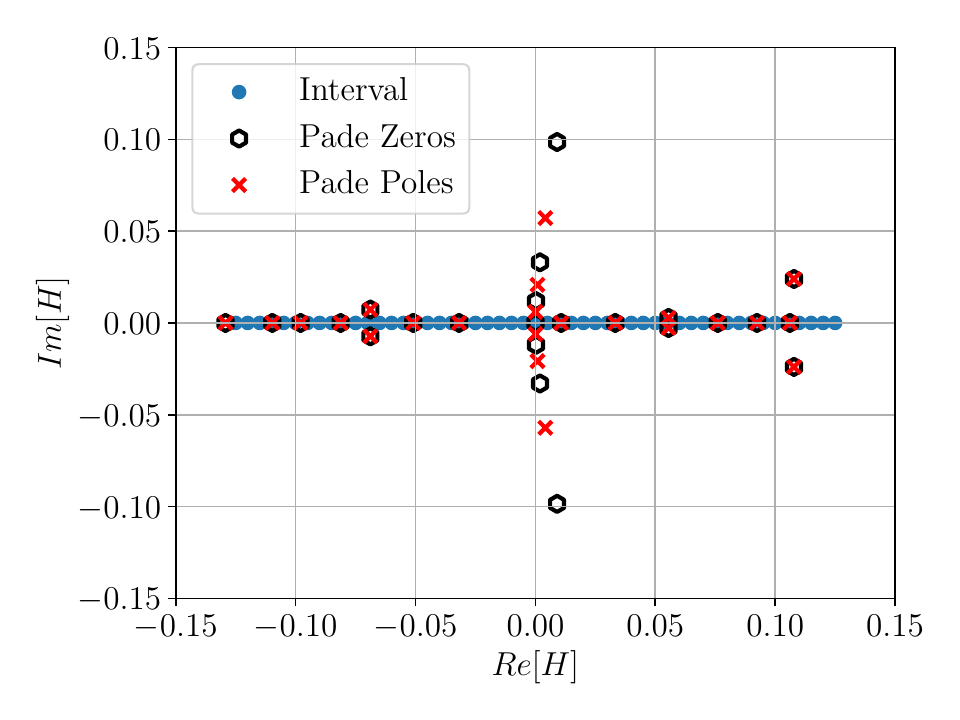}
	\caption{Zeros of
          the numerator (black pentagons) and of the denominator (red
          crosses) of the rational approximant $R^{m}_{n}(H)$ for the
          magnetisation on $L=15$ (Top) and $L=30$ (Bottom), with $[m,n] = [25,25]$. The pale blue circles are the points used as input for the Pad\'{e}. Notice that the closest singularity to the real axis gets closer to the real $H$ axis as $L$  gets larger, with real parts being consistent with $\mbox{Re}(H_0) = 0$.}
        \label{fig:LeeYangZerosL}
\end{figure}

\begin{figure}
    \centering
    \includegraphics[width=.52\textwidth]{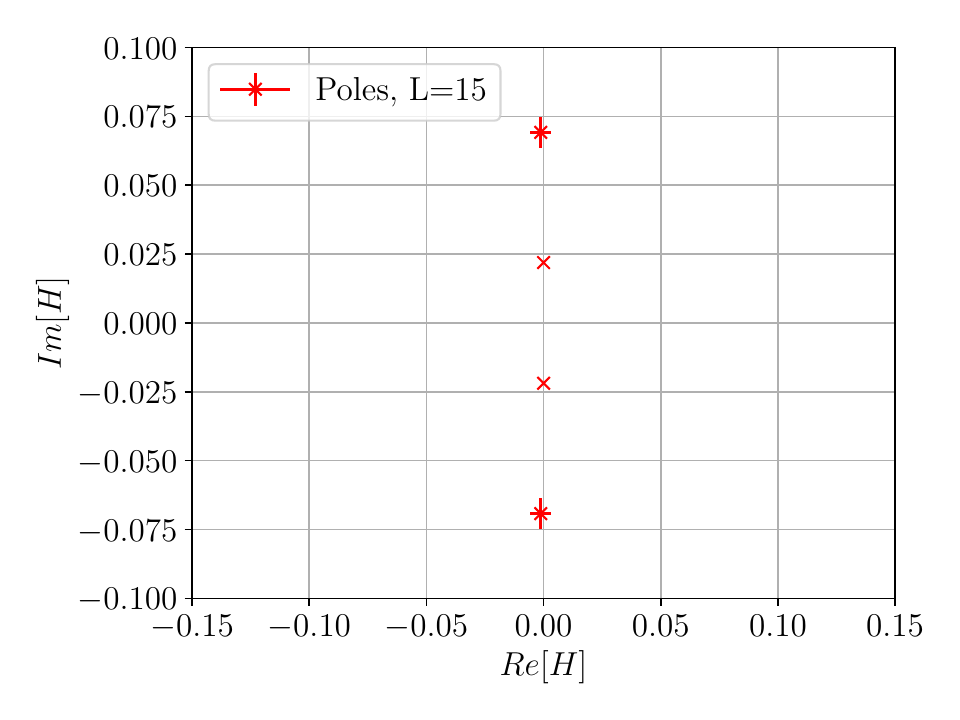}
    \hfill
    \includegraphics[width=.52\textwidth]{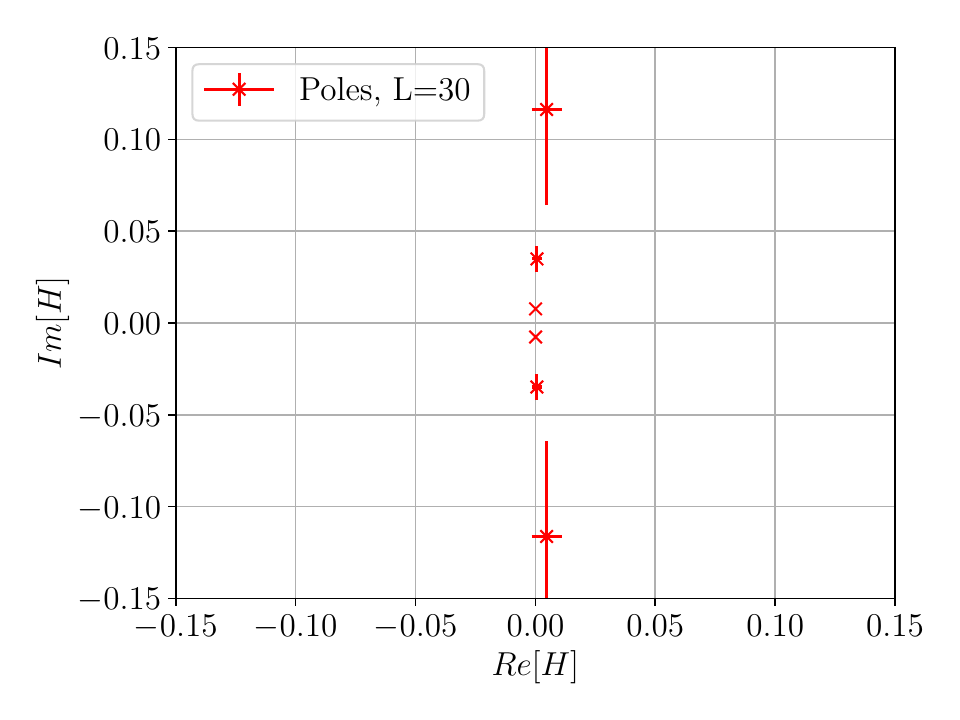}
    
    \caption{Genuine poles extracted from Fig.~\ref{fig:LeeYangZerosL} along with their error bars, shown for two lattice sizes (Top : $L = 15$, Bottom : $L = 30$). We want to highlight that these poles follow the \emph{circle theorem}. They lie on the imaginary $H$ axis (within errors) and the closest pole moves closer to the real $H$ axis as the lattice size increases. Moreover, the number of poles increases. }
    \label{fig:cir_theorem}
\end{figure}

\begin{figure}
    \centering
    \includegraphics[width=.52\textwidth]{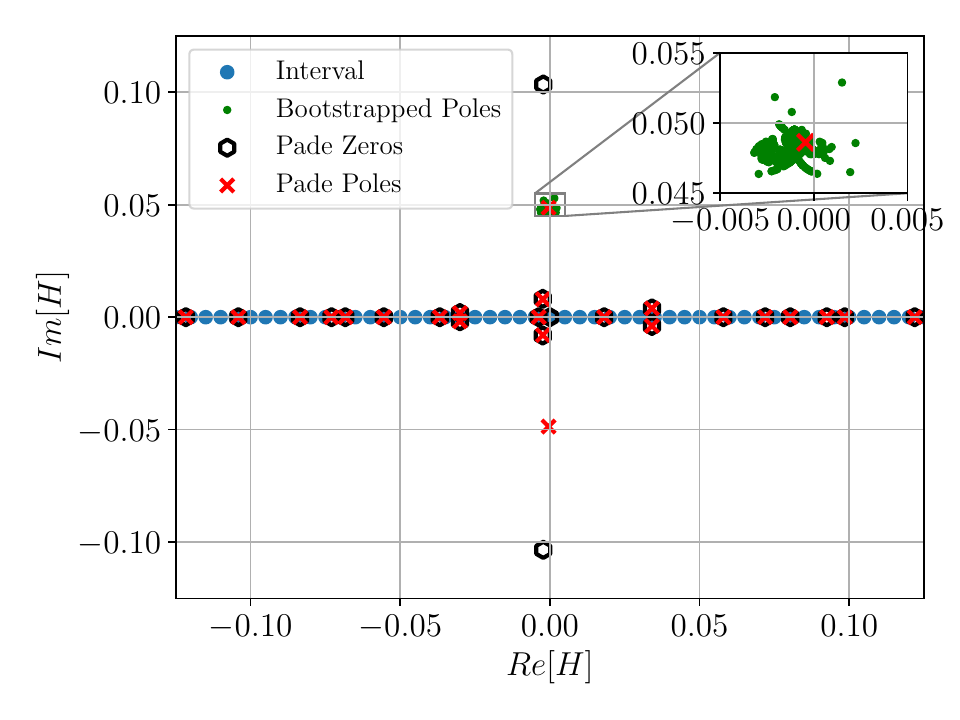}
    \caption{Stability of closest pole for $L = 10$. The green cloud represents closest poles extracted from varying the Taylor coefficients with noise drawn from a Gaussian distribution. The singularity structure shown, red crosses and black pentagons, are for the central values of the data at $L=10$. This is the result of drawing the coefficients $\sim 700$ times using a rational function of order $[m,n]=[25,25]$}
    \label{fig:sing_scatter}
\end{figure}

\begin{figure}
    \centering
    \includegraphics[width=.52\textwidth]{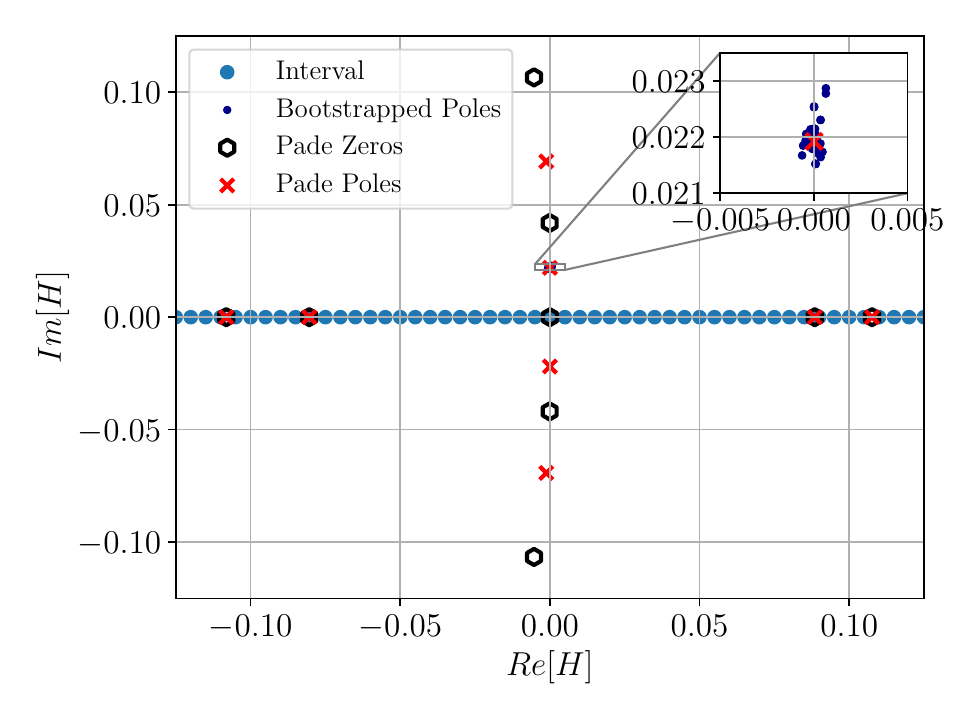}
    \caption{Stability of closest pole for $L = 15$. The dark blue cloud of points represents closest pole extracted by varying the order of the input Taylor coefficients using only the central values. The orders of the rational approximation used to obtain the blue cloud has been varied between  $[5,5]$ to $[25,25]$. }
    \label{fig:sing_scatterSYS}
\end{figure}

\noindent Before moving on to discuss the scaling of these zeros, which will make use of the closest poles extracted for each lattice size, we would like to briefly discuss the procedure we have used to attach error bars on the locations of the poles in the complex plane of $H$ and $\beta$. An important point to make is that in the absence of noise in data ({\em e.g.} one can construct this by discretizing a known function), if there is a genuine singularity of the function, it will appear as a stable pole of the rational approximation constructed. However, if the data is noisy, as it always is when dealing with simulation results, even if there is a genuine singularity of the function, the resulting pole will move, commensurate with the amount of noise present. Here, we can distinguish between two types of errors the poles can have, although they are not strictly independent.
\begin{itemize}
    \item \textbf{Statistical errors}: These are the errors propagated from the estimated error on the measured Taylor coefficients to the poles of the rational function approximant. The procedure used to estimate this error was to solve the system of linear equations in Eq.~(\ref{eq:LinearProblem3}) repeatedly by choosing new Taylor coefficients for each solve. These coefficients are drawn from a Gaussian distribution centered at the central values of the measured coefficients and having standard deviation given by the estimated error on the corresponding Taylor coefficient. We refer the reader to Fig.~\ref{fig:sing_scatter}, where the cloud of green points are the closest poles extracted for the $L=10$ lattice. The scatter is from repeating the bootstrap procedure around $\sim700$ times, keeping the order of the Pad\'e approximant fixed at $[m,n] = [25,25]$.
    \item \textbf{Systematic errors}: These are the errors on the closest poles resulting from varying the order of the multi-point Pad\'e and (or) changing the selection of the input points used to construct the Pad\'e approximant, using only the central values of the input data. The idea is to change the input points by deleting data in a systematic way to construct the rational function of varying orders. We vary the order of the Pad\'e approximant from $[m,n] = [25,25] \to [5,5]$. We now refer to Fig.~\ref{fig:sing_scatterSYS} where we show the singularity structure for $L=15$ lattice \footnote{The reader may notice this is a slightly different plot than that of Fig.~\ref{fig:LeeYangZerosL} (Top), although they are both using the same input data for the $L=15$ lattice (all available Taylor coefficients at their central values). The reason is that the rational functions have both been calculated using different methods : the one in Fig.~\ref{fig:LeeYangZerosL} using the Linear solver and the one in this plot using the least square fitting procedure. The goal is to further show the stability of the genuine structure of the poles obtained}. Here the scatter of poles arising from changing the order of the rational function is shown as dark blue points, notice that the scatter is over only around $\sim50$ points, as the goal is only to show the stability of the closest pole.
\end{itemize}

\noindent Note that the systematic errors mentioned above are correlated with the statistical errors. All in all, from the error analysis performed above, the stability of the closest poles of average magnetization gives us confidence in the fact that we are extracting genuine poles of the function, \emph{i.e.} genuine zeros of the partition function and thus LY zeros. We can now proceed to analyse the scaling of these poles with the lattice volume to extract physical information like the critical exponents.

\section{\label{sec:ScalingLY}Scaling analysis of zeros}

\noindent Until now we have mainly focused on partition function zeros arising in the complex $H$ plane (LY zeros) when considering cumulants at fixed temperature and varying $H$. However, looking at Eq.~(\ref{eqn:partn}) we can also consider the partition function zeros in the complex inverse temperature ($\beta$) plane. This has been done numerically in \cite{Deger:2019mgo}, where the authors have studied the Fisher and Yang-Lee zeros of the
2D and 3D Ising model by using a relatively high number of cumulants in the temperature and external magnetic field variables. As explained before, instead of using such high order of cumulants, we have made use of the multi-point Pad\'{e} method to study only two different cumulants
as a function of temperature and external magnetic field. We refer again to the Hamiltonian of Eq.~(\ref{eqn:hamil}), in which we set $J$ to unity. To draw parallel with the analysis in \cite{Deger:2019mgo}, we expand the partition function in terms
of its zeros in the $\beta$ plane, 
\begin{equation}
Z(\beta,H) = Z(0,H) \, e^{c \,\beta } \, \prod_k \left(1-\frac{\beta}{\beta_k}\right)
\end{equation}
$c$ being some constant and the product is over the $k$ zeros given by $\{\beta_k \}$. Thermal cumulants are defined by the relation
$$ \left\langle\langle U^n \right\rangle\rangle =
\frac{\partial^n}{\partial (-\beta)^n} \ln Z(\beta,H)$$
which using the expansion above can be re-expressed as,
\begin{equation}
\left\langle\langle U^n \right\rangle\rangle =
(-1)^{(n-1)} \sum_k \frac{(n-1)!}{(\beta_k-\beta)^n} \;\;\;\;\;\;\; (n>1)
\label{eq:finiteVol}
\end{equation}
Looking at Eq.~(\ref{eq:finiteVol}) above, it is easy to see that near criticality, the closest zero to the real axis will contribute the most to the thermal cumulant. Additionally, it is possible to study the finite volume scaling of the Fisher zero following \cite{Itzykson:1983gb,Janke:2000ca,Janke:2001by}, and the relations describing the approach of leading zeros to critical inverse temperature can be written as
\begin{equation}
\label{eq:scalingZeros1}
 \mbox{Im}(\beta_0)  \propto L^{-1/\nu}
\end{equation}
and 
\begin{equation}
\label{eq:scalingZeros2}
| \beta_0-\beta_c | \propto L^{-1/\nu} 
\end{equation}

\noindent where $\beta_0$ is the Fisher zero, resulting in
the closest singularity of cumulants to the real axis \footnote{$\beta_0$ shows up together with
  its complex conjugate $\beta^*_0$.}, $\beta_c$ is the critical inverse
temperature and $\nu$ is the relevant critical exponent, which describes the divergence of the correlation length with respect to temperature, near criticality. The proportionality constants in Eq.~(\ref{eq:scalingZeros1}) \& (\ref{eq:scalingZeros2}) are related to the infinite volume scaling function for the energy density \cite{Deger:2019mgo}.\\

\subsection{Extracting $\nu$ and $\beta_c$}
\noindent In order to determine these critical quantities, we will follow the steps (which we have previously also described in \cite{DiRenzo:2023xeg}): {\em (1)} we compute the $n=2$
thermal cumulant ({\em i.e.} the specific heat) at
  various inverse temperatures $\beta$ and lattice sizes $L$; {\em (2)} for each $L$ we compute the rational approximant
  $R^{m}_{n}(\beta)$ by our multi-point Pad\'e method; {\em (3)} at each
  $L$ we find the Fisher zero $\beta_0$, which is obtained as the
  the closest singularity of the cumulant to the real axis; {\em (4)} we study the finite size scaling
  of the values of $\beta_0$. We refer the reader to the first row of Table~\ref{tab:ResultsFit} for the results for $\nu$.
\begin{figure}[ht]
	\centering
	\includegraphics[width=.495\textwidth]{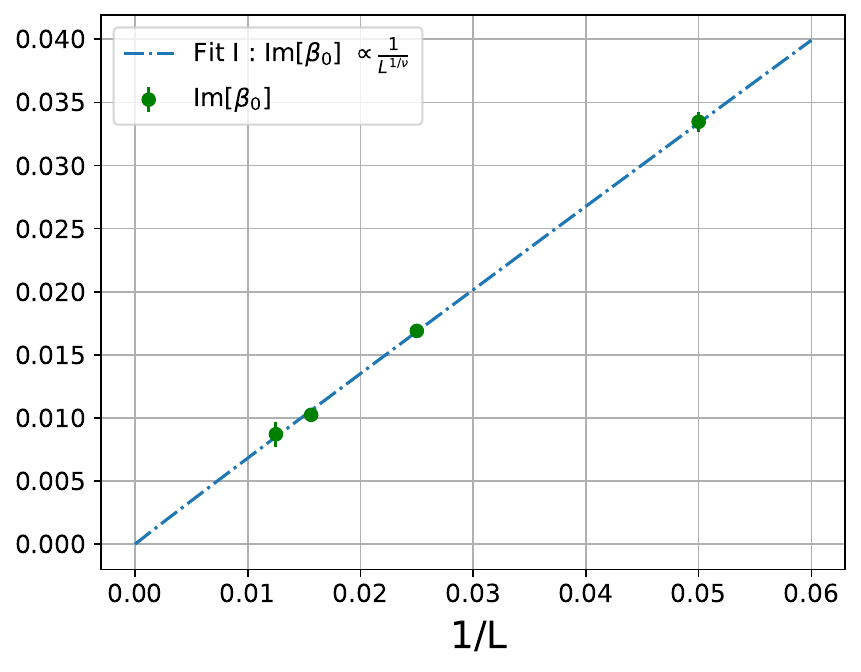}
	\hfill
	\includegraphics[width=.495\textwidth]{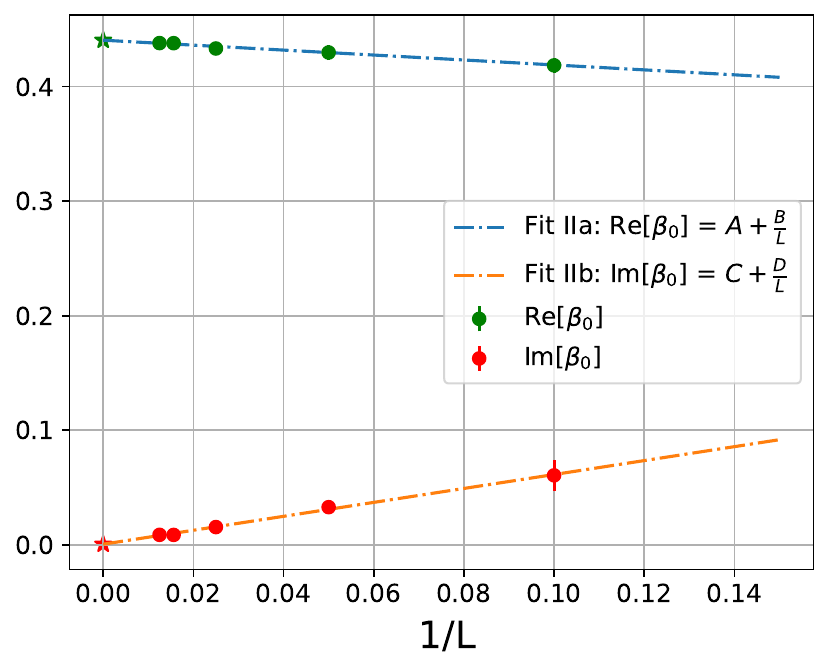}
	\caption{\label{fig:FisherZeros}(Top) The scaling in
          $1/L$ of $\mbox{Im}(\beta_0)$, {\em i.e.} 
          the imaginary part of the Fisher zero, detected as the
          closest singularity of the cumulant to
          the real axis. The correct critical exponent $\nu=1$ is reproduced
          with fairly good accuracy. (Bottom) Once $\nu$ has been
          extracted from the data, one can fit the value of the
          critical inverse temperature $\beta_c$ given by the intercept $A$ marked by a green star, which is 
          reconstructed to $1\%$ accuracy. The red star marks the intercept of Im$\beta_0$ as $L \to \infty$ which is recovered to be zero.}
\end{figure}
\begin{itemize}
\item Using Eq.~(\ref{eq:scalingZeros1}), we will try to extract the critical exponent $\nu$ using the following fit 
\begin{equation}
    \textrm{\textbf{Fit I}} :  \log{ \textrm{Im}[\beta_0] } = A + \frac{1}{B}*\log{\frac{1}{L}}
    \label{eq:fitI}
\end{equation}
with $A$ being the logarithm of the proportionality constant in Eq.~(\ref{eq:scalingZeros1}) and $B$ is the exponent of $\frac{1}{L}$ that we want to extract and compare its value with $\nu$.
As can be seen in Table.~\ref{tab:ResultsFit}, the value of the relevant critical exponent $\nu$ is obtained with decent accuracy with a value of $1.014(60)$, its exact value being $\nu=1$ \cite{Onsager1944}. Shown in the top panel of Fig.~\ref{fig:FisherZeros} is a pictorial description of the fit described in Eq.~(\ref{eq:fitI}). We plot $\mbox{Im}(\beta_0)$ as a function of $1/L$. On account of $B\sim1$, the dash-dotted line (which is the result of the fit) can hardly be distinguished from a straight line.  
\item Using Eqs.~(\ref{eq:scalingZeros1}) and (\ref{eq:scalingZeros2}) and the fact that the exponent $\nu = 1$, it is not hard to see that one can obtain a linear relation for Re[$\beta_0$] as a function of $1/L$ and define a fit function to extract $\beta_c$ as follows
\begin{equation}
    \textrm{\textbf{Fit IIa}} :  \textrm{Re}[\beta_0]  = A + B*\frac{1}{L}
    \label{eq:FitIIa}
\end{equation}
\begin{equation}
    \textrm{\textbf{Fit IIb}} :  \textrm{Im}[\beta_0]  = C + D*\frac{1}{L}
    \label{eq:FitIIb}
\end{equation}
Here $\nu = 1$ simplifies Eq.~(\ref{eq:scalingZeros2}) with $\textrm{Re}[\beta_0] \to \beta_c$ in the limit $L \to \infty$. Hence, in order to determine $\beta_c$, we will fit the real part of the closest Fisher zeros to the real $\beta$ axis as a function of $1/L$ and extract the intercept $A$. This intercept is shown as a green star in the bottom panel of Fig.~\ref{fig:FisherZeros}. Also for this our estimate seems fairly accurate at 
  $\beta_c=0.4404(19)$, when compared with the exact result of $\beta_c \sim 0.4407$ \cite{Onsager1944}. We additionally show in the same figure, that after identifying the exponent, one can also find the intercept of the line $\mbox{Im}(\beta_0)$ vs $1/L$ and show that it goes to zero within errors as seen with the red star on the figure.

\end{itemize}

\subsection{Extracting $\beta \delta$}
\begin{figure}[t]
\centering
\includegraphics[width=.495\textwidth]{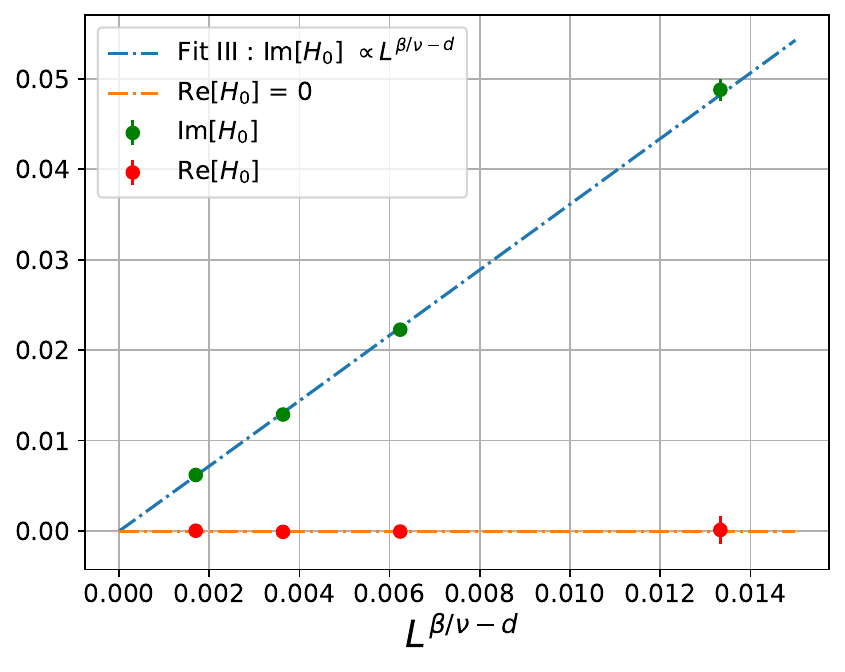}
\caption{Finite size scaling of $\mbox{Im}(H_0)$. To guide the eye, we
plot data versus $L^{1/8-2}$, where the correct critical exponents $\beta$ and $\nu$ are
taken. The value obtained from fits is $ \beta/\nu - d = -\beta \delta = -1.881(93)$, as shown in Table~\ref{tab:ResultsFit}, which also gives $\beta = 0.119(93)$. }
\label{fig:LeeYangScaling}
\end{figure}

\noindent After obtaining the inverse critical temperature, we can now perform simulations at $\beta_c$, to study the closest zero $\mbox{Im}(H_0)$ to the real axis in the complex $H$ plane as a function of lattice volume $L$. This has been the focus of most of the previous discussions in Sections~\ref{sec:LYZ}~\&~\ref{sec:Pade}. Once again following the procedure outlined in \cite{DiRenzo:2023xeg},  
our program again entails the following steps: {\em (1)} we compute the $n=1$
magnetic cumulant ({\em i.e.} the magnetisation) at
  $\beta=\beta_c$ and various values of external magnetic field $H$ and 
  lattice size $L$; {\em (2)} for each $L$ we compute the rational approximant
  $R^{m}_{n}(H)$ for the magnetisation by our multi-point Pad\'{e} method; {\em (3)} at each $L$
  we find the Lee Yang zero $H_0$, which is the singularity of the
  rational approximant for the  magnetisation which is the closest 
to the real axis; {\em (4)} we study the finite size scaling
  of the values of $\mbox{Im}(H_0)$ (as we have seen in Fig. \ref{fig:cir_theorem} , $H_0$ always sits
  at $\mbox{Re}(H_0)=0$), given by \cite{Itzykson:1983gb,Deger:2019mgo}:
  
  \begin{align}
      | H_0 - H_c | &\propto L^{\beta/\nu - d} 
     \label{eq:scalLY}
  \end{align}

\noindent where, the exponent $\beta$ is the well known critical exponent that describes how the average magnetization goes to zero when we approach the critical point from below $T_c$ and $d$ is the dimension. The proportionality constant in Eq.~(\ref{eq:scalLY}) is related to the infinite volume scaling function for the total magnetization. In order to extract the exponent in Eq.~(\ref{eq:scalLY}), we will fit the following function
\begin{equation}
    \textrm{\textbf{Fit III}} :  \log{ \textrm{Im}[H_0] } = A + B*\log{L}
    \label{eq:FitIII}
\end{equation}
with $A$ being the logarithm of the proportionality constant in Eq.~(\ref{eq:scalLY}) and $B$ the exponent of $L$ that we want to extract and compare with $\beta/\nu - d$. Using the known scaling relations between the standard critical exponents, we can derive the following hyperscaling relation between $\beta,\nu,d$ and $\delta$,  
\begin{align}
    \nu d &= \beta \, (1+\delta) \nonumber \\
    \implies \frac{\beta}{\nu} - d &= -\frac{\beta \delta}{\nu}
    \label{eq:hypscalrel}
\end{align}
\noindent Remembering the value of $\nu=1$ for the 2D Ising model, we can thus use the fit result for the parameter $B$ to estimate $\beta \delta$. As can be seen from Table~\ref{tab:ResultsFit}, the fit parameter $ B = -1.881(93)$ has been determined, which gives to decent precision the estimate for $\beta \delta$ whose exact value for the 2D Ising model is $1.875$. Further, without using the hyperscaling relation given in Eq.~(\ref{eq:hypscalrel}), the fit parameter $B$ should be compared against $\beta/\nu - d$, to obtain $\beta = 0.119(93)$ which compared against its exact value of $0.125$ for the 2D Ising model, gives an estimate for the exponent.  
Finally, we show the results of the fit of $H_0$ we obtained for each lattice size, plotted against $L^{\beta/\nu - d}$ in Fig.~\ref{fig:LeeYangScaling}. 
In principle one should be able to follow these steps to estimate the critical region for QCD using Taylor expansions from lattice QCD. The relevant parameters which control the critical region in QCD will be the baryon density. We have tried to make some concrete steps in this direction recently in \cite{DiRenzo:2024izy}. However, being a much more complicated theory, both numerically and conceptually, we may have to wait for some time to be able to do that. 

\begin{table}[!ht]
    \centering
    \resizebox{0.5\textwidth}{!}{
    \begin{tabular}{|c|c|c|c|c|} \hline
         \textbf{FIT} & $A$  & $B$ & exact  & $\chi^2/dof$  \\ \hline
         {I}&
         {$-0.446(209)$}&
         {$\mathbf{1.014(60)}$}&
         {$\nu = 1$}&
         {$1.3$}\\
         {IIa}&
         {$\mathbf{0.4404 (19)}$}&
         {$-0.216 (70)$}&
         {$\beta_c \sim 0.4407$}&
         {$1.44$} \\ 
         {III}&
         {$1.30 (24)$}&
         {$\mathbf{1.881 (93)}$}&
         {$d - \beta/\nu = \beta \delta = 1.875$}&
         {$1.2$}\\ \hline
    \end{tabular} }
    \caption{Results for the fits shown in Eqs.~(\ref{eq:fitI},\ref{eq:FitIIa},\ref{eq:FitIII}), shown for each row respectively. For Fit I, $B$ has to be compared with the exact value stated, whereas for Fit II, the intercept gives $\beta_c$, hence the exact value has to be compared with $A$. For the last fit, Fit III, the fit parameter $B$ has to be compared to the exact value of the critical exponent product, namely, $\beta \delta$. }
    \label{tab:ResultsFit}
\end{table}

\section{\label{sec:Conclusions}Conclusions and Outlook}

\noindent As a first step we simulated the 2D Ising model using a cluster spin flip algorithm in two ways. For the LY zero analysis, we simulated the model on varying lattice sizes at a set of values of the external magnetic field. These simulations were performed at $T_c$, and the goal was to approximate the average magnetization as a rational function of the external magnetic field and study the structure of zeros and poles that arise. On the one hand we were able to verify numerically, many properties of the LY zeros including the famous circle theorem for the Ising model, observing that only the genuine poles (un-cancelled and stable) of magnetization lie on the purely imaginary $H$ axis. It was further observed that the number of genuine poles poles increases with volume and for simulations at $T_c$, comes closer to the real $H$ axis. In order to verify that these were indeed physical effects, volume scaling of the zeros, using the prescription in \cite{Itzykson:1983gb,Deger:2019mgo} were performed leading to a decent estimate of the combination of critical exponents $\beta \delta = 1.881(93)$. The fact that LY zeros can be studied at $T_c$ is in our approach fully self-consistent. In fact, Fisher zeros were also studied by approximating the specific heat with a multi-point Pad\'{e} function and studying its poles in the complex $\beta$ plane. Finite size scaling of these zeros following the prescription of \cite{Deger:2019mgo} was done to obtain again, precise values of the critical exponent $\nu = 1.014(60)$ and of the critical inverse temperature $\beta_c = 0.4404(19)$. These results give us some confidence in our pursuits of studying the QCD phase diagram using lattice QCD simulations combined with multi-point Pad\'{e} method. The main caveat being that in the case of the Ising model it was not very computationally expensive to reach statistics of the order of $\sim 625$K configurations for each lattice, for each value of $H$ and $\beta$. These kind of statistics are not currently realistic for lattice QCD simulations. 

\noindent Another direction to pursue would be to study the universal location of the Lee-Yang edge singularities as done using the Functional Renormalisation Group (FRG) approach in \cite{AConneliPRL2019LY,GJ:LYE2023thesis} or using a suitable parametrization of lattice data as done in \cite{karsch2023leeyang} for the $O(N)$ models. In order to do this using our approach, we would need accurate values of the closest LY zero to the real $H$ axis, but at temperatures $T > T_c$. We would like to close by inviting the reader to use the multi-point Pad\'{e} method for extracting Lee-Yang and Fisher zeros in their choice of models. All data from our calculations, presented in the ﬁgures of this paper, can be found in \cite{singh2024dataset}.

\section*{Acknowledgements}
S.S. would like to thank Christian Schmidt, Frithjof Karsch and Vladimir Skokov for interesting discussions on the critical phenomena and Lee-Yang zeros.
This work was supported by the (i) European Union’s Horizon 2020 research and innovation program under the Marie Sklodowska-Curie Grant Agreement No. H2020-MSCAITN-2018-813942 (EuroPLEx), (ii) Part of this work was done with S.S. being supported by PUNCH4NFDI consortium
supported by the Deutsche Forschungsgemeinschaft (DFG, German Research Foundation) – project number 460248186 (PUNCH4NFDI), (iii) I.N.F.N. under the research project i.e. QCDLAT. A good fraction of this research used computing resources made available through the University of Parma on the UNIPR HPC facility. 

\bibliography{Ising2D}

\end{document}